\documentclass[twocolumn,showpacs,preprintnumbers,amsmath,amssymb]{revtex4}

\usepackage{graphicx}
\usepackage{dcolumn}
\usepackage{bm}

\begin{document}

\preprint{}

\title{Fermi Surface Reconstruction without Breakdown of Kondo Screening\\at Quantum Critical Point}

\author{Hiroshi Watanabe}
 \email{hwatanabe@hosi.phys.s.u-tokyo.ac.jp}
\author{Masao Ogata}%
\affiliation{%
Department of Physics, University of Tokyo, Hongo, Bunkyo-ku, Tokyo 113-0033, Japan
}%

\date{\today}

\begin{abstract}
Motivated by recent Hall-effect experiment in YbRh$_2$Si$_2$, we study ground state properties of a Kondo lattice model 
in a two-dimensional square lattice using variational Monte Carlo method. We show that there are two types of phase transition,
antiferromagnetic transition and topological one (Fermi surface reconstruction). In a wide region of parameters, these two 
transitions occur simultaneously without the breakdown of Kondo screening, accompanied by a discontinuous change of the Hall
coefficient. This result is consistent with the experiment and gives a novel theoretical picture for the quantum critical point 
in heavy fermion systems.
\end{abstract}

\maketitle
Quantum phase transition where a transition temperature goes to zero is a very interesting phenomena often observed in 
strongly correlated electron systems and has attracted much interests these days.
Compared with the conventional finite-temperature phase transition, quantum nature of fluctuation explicitly appears.  
In heavy fermion systems, the quantum phase transition occurs when the antiferromagnetic (AF) long-range order disappears,
as some parameters such as pressure, doping and applied magnetic field changes. 
Around this quantum critical point (QCP), many interesting phenomena such as non-Fermi-liquid behavior or unconventional 
superconductivity have been observed. Recent experimental development makes it possible to study the property of the QCP in 
detail and suggests the necessity of the improvement of a conventional view for the QCP.

Recently, it was found that the Hall coefficient of a heavy fermion material, YbRh$_2$Si$_2$, shows a rapid change as a function of 
magnetic field~\cite{Paschen}. When the temperature is lowered, this crossover point approaches the QCP and simultaneously the crossover becomes
much sharper. From this result, it was claimed that the crossover becomes a discontinuous jump at $T=0$ and that a reconstrunction of the Fermi surface
(FS) occurs at the QCP, i.e., from a large FS to a small FS~\cite{Paschen}.
In the conventional view of heavy fermion systems, the conduction electrons ($c$-electrons) and the localized spins ($f$-electrons) 
hybridize with each other through the Kondo coupling and form ``heavy Fermi liquid" on both sides of the QCP.
In this case, the volume of the FS is determined by the total number of the $c$- and $f$-electrons and thus the system has a ``large FS".
Once such a Fermi liquid is formed, the self-consistent-renormalization (SCR) theory by Moriya~\cite{Moriya} and renormalization-group studies by
Hertz~\cite{Hertz} and Millis~\cite{Millis} can be applied. 
However, the FS reconstrunction observed experimentally can not be understood. Recently Si proposed an alternative theory for the QCP, 
called  as ``local quantum criticality"~\cite{Si1,Si2}.
He claimed that the Kondo screening becomes irrelevant at the QCP and the $c$- and $f$-electrons are decoupled in the AF phase. 
As a result, $f$-electrons turn to be localized and do not contribute to the FS any more. If this is the case, the volume of the FS is determined 
only by the number of $c$-electrons and a so-called ``small FS" is realized. Although this picture of local quantum criticality seems reasonable to
explain the Hall-effect experiment, it is still sketchy and the details of the theory are controversial.

In this paper, we study this problem microscopically by applying a variational Monte Carlo (VMC) method to the Kondo lattice model (KLM).
Since the Hall-effect experiment suggests a zero-temperature phase transition at QCP, the $T=0$ variational approach is suitable to this problem.
The VMC method has been successfully used in the strongly correlated electron systems. 
We construct several trial wave functions which have large- or small FS together with AF long-range order. Evaluating the variational energies with 
VMC simulation, we determine the ground state phase diagram as a function of the Kondo coupling $J$ and the $c$-electron number $n_{\mathrm{c}}$.
We find a first order phase transition from a ``large FS" to a ``small FS" as a function of $J$, even if the Kondo screening always occurs.
The present results give a new route of the FS reconstrunction which was not considered in local quantum criticality.

\begin{figure*}
\includegraphics[width=17cm]{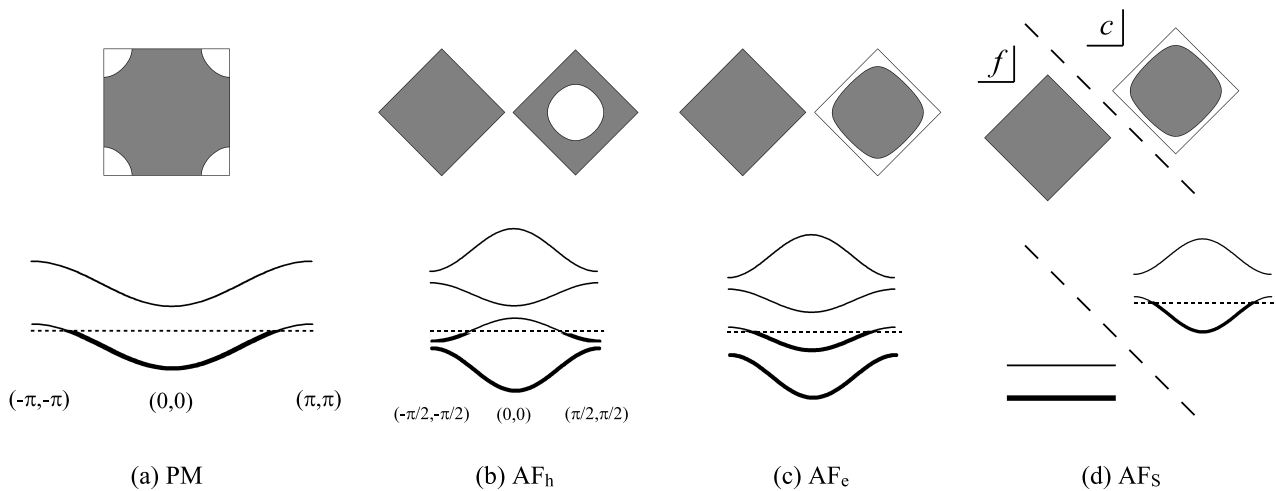}
\caption{Fermi surfaces and band dispersions of the four different variational states considered in this paper.
(a) paramagnetic metal (PM), (b) AF$_{\mathrm{h}}$, (c) AF$_{\mathrm{e}}$ and (d) AF$_{\mathrm{S}}$.}
\label{fig1}
\end{figure*}

We study the following KLM in a two-dimensional square lattice,
\begin{equation}
 H=\sum_{\bm{k}\sigma}\epsilon_{\bm{k}}c^{\dagger}_{\bm{k}\sigma}c_{\bm{k}\sigma}
  +J\sum_{i}\bm{S}_i\cdot\bm{s}_i \label{KLM}
\end{equation}
where $\bm{S}_i$ and $\bm{s}_i$ represent the $f$- and $c$-electron spins, namely, 
$\bm{S}_i=\frac{1}{2}\sum_{\sigma\sigma'}f^{\dagger}_{i\sigma}\bm{\sigma}_{\sigma\sigma'}f_{i\sigma'}$ and 
$\bm{s}_i=\frac{1}{2}\sum_{\sigma\sigma'}c^{\dagger}_{i\sigma}\bm{\sigma}_{\sigma\sigma'}c_{i\sigma'}$.
$J(>0)$ denotes the AF exchange coupling between them.
As for the kinetic energy term, we consider the nearest-neighbor hopping and thus the bare $c$-electron energy dispersion is given by
\begin{equation}
 \epsilon_{\bm{k}}=-2t(\cos k_x+\cos k_y). \label{Ek}
\end{equation}
In principle, both the Kondo effect and the RKKY interaction are the consequences of the ``intrasite" interaction between 
$c$- and $f$-electron spins.  
However, many mean-field type theoretical studies include an additional term of the ``intersite" exchange interaction,
$J_{\mathrm{H}}\sum\bm{S}_i\cdot\bm{S}_j$, for theoretical convenience (This model is called ``Kondo Heisenberg model"). 
We have to deal with this additional term with care, since it sometimes leads to overestimate of the RKKY interaction 
and even to unphysical results.
Therefore, we study the model without the intersite exchange term in this letter.

The KLM has been studied in a mean-field approximation (MFA)~\cite{Zhang,Senthil}.
The disadvantage of MFA is that the local constraint of one electron per $f$ orbital, $\sum_{\sigma}f^{\dagger}_{i\sigma}f_{i\sigma}=1$,
is not satisfied locally but only in average. 
Indeed, other methods which can correctly enforce this local constraint at $n_{\mathrm{c}}=1.0$ have shown that the MFA gives 
wrong results~\cite{Capponi,Jurecka}.
In contrast, the VMC method takes account of the local constraint exactly. In the following, we focus on the case with 
$n_{\mathrm{c}}<1.0$ where the issue of large- and small FS can be discussed. 
For this case ($n_{\mathrm{c}}<1.0$), the conventional Quantum Monte Carlo method~\cite{Capponi} can not be used due to the severe sign problem.
The advantage of using the VMC method is twofold: (1) it has no negative sign problem, 
and thus can be applied to the system with $n_{\mathrm{c}}<1.0$ or with geometrical frustration, and (2) the local constraint of each site is  
strictly satisfied. This constraint is necessary for the study of the KLM, but hard to be enforced by other methods. 

For the trial wave functions, we use the following Jastrow-type wave functions,
\begin{equation}
 \left|\Psi \right> = P_{n^f_{\mathstrut}=1} \left|\Phi \right>,
\end{equation}
where
\begin{equation}
 P_{n^f_{\mathstrut}=1} = \prod_{i}\left[ n^f_{i\uparrow}(1-n^f_{i\downarrow})+n^f_{i\downarrow}(1-n^f_{i\uparrow})\right],
\end{equation}
is a projection operator which keeps the $f$-electron number of each site exactly one.
$\left|\Phi \right>$ is obtained by diagonalizing a one-body Hamiltonian with some variational parameters.
This type of wave function has been studied by Shiba and Fazekas for the one-dimensional case~\cite{Shiba}.
However, AF long-range order is not stabilized in one-dimension, so that the AF QCP was not studied~\cite{Tsunetsugu,Shibata}.
In contrast, we show that the AF state is stabilized in a finite region of $J<J_{\mathrm{c}}$ and we can discuss the AF QCP.

First, let us explain the trial wave functions used in this paper in some detail.
For a paramagnetic metal (PM) state, we construct $\left|\Phi \right>$ from the one-body Hamiltonian
\begin{equation}
H=\sum_{\bm{k}\sigma}
    \left(c_{\bm{k}\sigma}^{\dagger}, f_{\bm{k}\sigma}^{\dagger}\right)
  \begin{pmatrix}
   \epsilon_{\bm{k}} & -\tilde{V} \\ -\tilde{V} & \tilde{E_f}
 \end{pmatrix}
 \begin{pmatrix}
   c_{\bm{k}\sigma} \\ f_{\bm{k}\sigma}
 \end{pmatrix} \label{PM},
\end{equation}
where $f_{\bm{k}\sigma}$ is a Fourier transform of $f_{i\sigma}$.
$\tilde{V}$ and $\tilde{E_f}$ are variational parameters that control the degree of $c$-$f$ hybridization 
and the effective $f$-electron level, respectively.
Since the ratios $\tilde{V}/t$ and $\tilde{E_f}/t$ appear in the trial wave function, we do not need to treat $t$ in 
$\epsilon_{\bm{k}}$ as a variational parameter. We set $t=1$ as a unit of energy in the following.

In the same way, AF state can be obtained from
\begin{align}
H&=\sum_{\bm{k}\sigma}
    \left( c_{\mathrm{A}\bm{k}\sigma}^{\dagger}, c_{\mathrm{B}\bm{k}\sigma}^{\dagger},  
   f_{\mathrm{A}\bm{k}\sigma}^{\dagger}, f_{\mathrm{B}\bm{k}\sigma}^{\dagger}\right) \times \notag \\
  &\begin{pmatrix}
   \sigma m & \epsilon_{\bm{k}} & -\tilde{V} & 0 \\ 
   \epsilon_{\bm{k}} & -\sigma m & 0 & -\tilde{V} \\
   -\tilde{V} & 0 & \tilde{E_f}-\sigma M & 0 \\
   0 & -\tilde{V} & 0 & \tilde{E_f}+\sigma M 
 \end{pmatrix}
 \begin{pmatrix}
  c_{\mathrm{A}\bm{k}\sigma} \\ c_{\mathrm{B}\bm{k}\sigma} \\ 
  f_{\mathrm{A}\bm{k}\sigma} \\ f_{\mathrm{B}\bm{k}\sigma}    
 \end{pmatrix}, \label{AF}
\end{align} 
where $m$ and $M$ denote the variational parameters of AF moment in $c$- and $f$-electrons, respectively.
A and B represent the indices of each sublattice. The summation over $\bm{k}$ is in the folded AF Brillouin zone.
If we set $m=M=0$, Eq.~(\ref{AF}) reduces to Eq.~(\ref{PM}).

We optimize the variational parameters mentioned above and find the lowest energy state in a $J-n_{\mathrm{c}}$ phase diagram.
Possible candidates for the ground state are classified according to the shape of the FS and the band dispersion, 
as shown in Fig.~\ref{fig1}: (a) paramagnetic metal (PM), 
(b) AF metal with hole-like FS (AF$_{\mathrm{h}}$), (c) AF metal with electron-like FS (AF$_{\mathrm{e}}$)
and (d) AF metal without $c$-$f$ hybridization (AF$_{\mathrm{S}}$, S denotes ``small"). 
In PM, AF$_{\mathrm{h}}$ and AF$_{\mathrm{e}}$ states, $\tilde{V}$ is finite, i.e., the $c$- and $f$-electrons  
hybridize with each other through the Kondo screening. 
The state called AF$_{\mathrm{S}}$ represents the small FS proposed by Si, which has no Kondo screening, $\tilde{V}=0$.
The state PM is the conventional Fermi liquid state with large FS, where quasiparticles become heavy due to the hybridization and
projection operator $P_{n^f_{\mathstrut}}=1$. When AF long-range order is introduced into PM, we obtain AF$_{\mathrm{h}}$.
Therefore, PM and AF$_{\mathrm{h}}$ are smoothly connected with each other and their FSs have the same topology (hole-like FS). 
On the other hand, AF$_{\mathrm{e}}$ (Fig.~\ref{fig1}(c)) is a state which can be obtained by introducing $\tilde{V}$ in the 
AF$_{\mathrm{S}}$ with small FS. This state have not been obtained before in the MFA, but it is a natural extension of AF$_{\mathrm{S}}$.
The topology of FS of AF$_{\mathrm{e}}$ is different from that of AF$_{\mathrm{h}}$ as shown in Fig.~\ref{fig1}(b) and (c).
However, since the Brillouin zone is halved due to the presence of AF long-range order, the FS volumes are the same.
Actually, by tuning $\tilde{V}$, $\tilde{E_f}$, $m$ and $M$, the band dispersion changes continuously from AF$_{\mathrm{h}}$ to
AF$_{\mathrm{e}}$ (Fig.~\ref{fig1}). In this sense, there is no difference of symmetry between AF$_{\mathrm{h}}$ and AF$_{\mathrm{e}}$,
but the Hall coefficient will be different between these states. We regard the state AF$_{\mathrm{h}}$ as a state with a
``large FS" connected to PM (Fig.~\ref{fig1}(a)), and AF$_{\mathrm{e}}$ as a state with ``small FS" connected to 
AF$_{\mathrm{S}}$ (Fig.~\ref{fig1}(d)). 

\begin{figure}
\includegraphics[width=7.0cm]{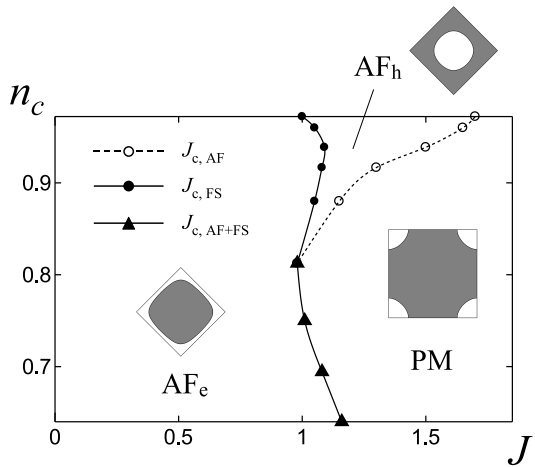}
\caption{Ground state phase diagram in the $J-n_{\mathrm{c}}$ plane obtained in VMC.
Solid lines ($J_{\mathrm{c,FS}}$ and $J_{\mathrm{c,AF+FS}}$) represent the first-order
phase transition and dotted line ($J_{\mathrm{c,AF}}$) represents the second-order one.}
\label{fig2}
\end{figure}

\begin{figure}
\includegraphics[width=6.5cm]{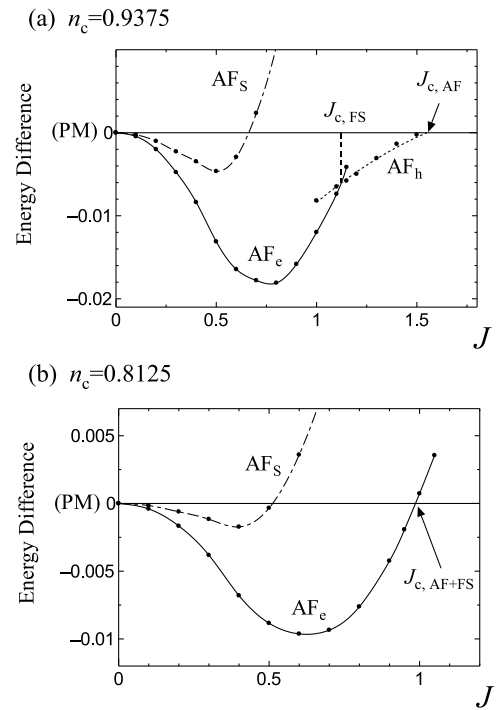}
\caption{Energy difference of each state compared with the PM state for (a) $n_{\mathrm{c}}=0.9375$ and (b) $n_{\mathrm{c}}=0.8125$.
The error bars are order of $10^{-4}$ and not shown here.}
\label{fig3}
\end{figure}

Figure~\ref{fig2} shows the obtained phase diagram in the $J-n_{\mathrm{c}}$ plane.
The numbers of sites are set to be 64, 100, 144 and 196 for the calculation and the results show good convergence.
For $n_{\mathrm{c}}\gtrsim 0.82$, we have two different types of transition. 
When $J$ decreases from a large value, the variational ground state changes from PM to AF$_{\mathrm{h}}$ at $J_{\mathrm{c,AF}}$. 
This is a ``conventional" second-order AF transition where the AF order parameter grows continuously from zero to a finite value. The first 
Brillouin zone is folded and the number of energy band doubles (2 $\rightarrow$ 4) as shown in Fig.~\ref{fig1}(a) to (b). 
The energy difference compared with the PM state is shown in Fig.~\ref{fig3}(a). 
When this transition occurs, the Hall coefficient must change continuously.  
Then, as $J$ decreases further, the state changes from AF$_{\mathrm{h}}$ to AF$_{\mathrm{e}}$ at $J_{\mathrm{c,FS}}$. 
We find that this transition is first-order even if AF$_{\mathrm{h}}$ and AF$_{\mathrm{e}}$ are connected in the variational parameter space.
Actually, the variational energy has a double minimum, and an energy crossing occurs as shown in Fig.~\ref{fig3}(a) at $J\simeq1.1$. 
In other words, the band dispersion changes discontinuously from convex downward to convex upward as shown in Fig.~\ref{fig1}(b) and (c).
It is regarded as a kind of Lifshitz transition, namely, the topology of FS changes. In this case, the discontinuous change of the Hall coefficient
occurs. It is a novel type of quantum phase transition originated from the competition between the Kondo effect and the RKKY interaction.
In addition to the Hall coefficient, the expectation value of each term in Eq.~(\ref{KLM}) shows a discontinuous change at $J_{\mathrm{c,FS}}$. 

\begin{table*}
\caption{Several pictures for the QCP in heavy Fermion material YbRh$_2$Si$_2$.}
\label{tab:table1}
 \begin{ruledtabular}
  \begin{tabular}{ccccc}
   Picture & Transition & Kondo screening & Change of Hall coefficient & Experiment \\ \hline
   conventional QCP & PM $\rightarrow$ AF$_\mathrm{h}$ & remains & continuous & inconsistent \\
   local quantum criticality (Si \textit{et al}.) & PM $\rightarrow$ AF$_\mathrm{S}$ & disappears & discontinuous & consistent \\
   our study $(n_{\mathrm{c}}\lesssim0.82)$ & PM $\rightarrow$ AF$_\mathrm{e}$ & remains & discontinuous & consistent \\  
  \end{tabular}
 \end{ruledtabular}
\end{table*}

When $n_{\mathrm{c}}\lesssim 0.82$, AF$_\mathrm{h}$ is not stabilized and the state changes directly from 
PM to AF$_{\mathrm{e}}$ at $J_{\mathrm{c,AF+FS}}$.
The obtained variational energies are shown in Fig.~\ref{fig3}(a).
In this case, the AF transition and the topological phase transition of FS occur simultaneously.
The AF order parameter and the Hall coefficient change discontinuously reflecting the character of the first order phase transition.

Interestingly, there is no region where AF$_{\mathrm{S}}$ is stable. Namely, the Kondo screening always occurs in our phase diagram.
This result is quite different from those obtained in MFA~\cite{Zhang,Senthil}, where the Kondo screening is suppressed across the QCP and 
the AF$_{\mathrm{S}}$ becomes stable in a wide region. 
We think that this discrepancy appears from the insufficient treatment of the local constraint in MFA. As discussed before for
the case with $n_{\mathrm{c}}=1.0$~\cite{Capponi,Jurecka}, the AF order is overestimated in MFA and thus the MFA is not appropriate for 
the problem considered here.   

Finally let us compare the Hall-effect experiment with the obtained results. 
We summarize the theoretical pictures for the QCP in Table~\ref{tab:table1}.
In a conventional picture, the Hall coefficient varies continuously
across the QCP~\cite{Coleman}. In our notation, the state changes from PM to AF$_{\mathrm{h}}$. 
In this case, even if there is a nesting of FS, the change is continuous unless the FS is perfectly flat~\cite{Norman}. 
Therefore, this conventional picture is inconsistent with the experiment.
On the other hand, we have found that the AF transition and the discontinuous change of the Hall coefficient occur simultaneously
for $n_{\mathrm{c}}\lesssim0.82$. 
This is consistent with the experiment, although it is not conclusive whether the Hall coefficient has a discontinuity at $T=0$ experimentally.
We would like to emphasize that our transition is different from the local quantum criticality by Si in the sense 
that the Kondo screening always occurs. We have shown that the FS topology changes discontinuously even in the presence of hybridization.
Our present results are $T=0$ variational calculation, and can not be applied directly to finite temperature.
Hall effect experiment was carried out at finite temperatures and only a crossover takes place as a function of applied magnetic field.
The application to finite temperature beyond MFA is a challenging problem in future.

In summary, we have studied the ground state of the KLM in a two-dimensional square lattice with the VMC method.
We proposed new types of wave functions (AF$_{\mathrm{h}}$ and AF$_{\mathrm{e}}$) which have different FS topologies.
It is shown that there are two kinds of phase transition, the AF transition and the topological one (FS reconstruction) even if 
the Kondo screening always occurs.
We find that these two transitions occur simultaneously at the QCP for $n_{\mathrm{c}}\lesssim0.82$ and this can explain the result of the 
Hall-effect experiment in YbRh$_2$Si$_2$. It is a novel type of quantum phase transition induced by the competition between the Kondo effect 
and the RKKY interaction.  

\begin{acknowledgments}
The authors thank K. Ueda, S. Nakatsuji and H. Harima for useful discussions. 
This work is supported by Grants-in-Aid from the Ministry of Education, Culture, Sports, Science and Technology of Japan and 
also by a Next Generation Supercomputing Project, Nanoscience Program, MEXT, Japan.
\end{acknowledgments}

\end{document}